# Insights from Device Modeling of Perovskite Solar Cells

*Nir Tessler[‡][*] and Yana Vaynzof[ᶠ][*]*

[‡] Sara and Moshe Zisapel Nano-Electronic Center, Department of Electrical Engineering, Technion-Israel Institute of Technology, Haifa 32000, Israel

[ᶠ] Integrated Centre for Applied Physics and Photonic Materials and Centre for Advancing Electronics Dresden (cfaed), Technical University of Dresden, Nöthnitzer Straße 61, 01187 Dresden, Germany

AUTHOR INFORMATION

**Corresponding Authors**

* E-mail: nir@technion.ac.il

* E-mail: yana.vaynzof@tu-dresden.de

ABSTRACT

In this perspective, we explore the insights into the device physics of perovskite solar cells gained from modeling and simulation of these devices. We discuss a range of factors that influence the modeling of perovskite solar cells, including the role of ions, dielectric constant, density of states, and spatial distribution of recombination losses. By focusing on the effect of non-ideal energetic alignment in perovskite photovoltaic devices, we demonstrate a unique feature in low recombination perovskite materials – the formation of an interfacial, primarily electronic, self-induced dipole that results in a significant increase in the built-in potential and device open-circuit voltage. Finally, we discuss the future directions of device modeling in the field of perovskite photovoltaics, describing some of the outstanding open questions in which device simulations can serve as a particularly powerful tool for future advancements in the field.

TOC GRAPHICS

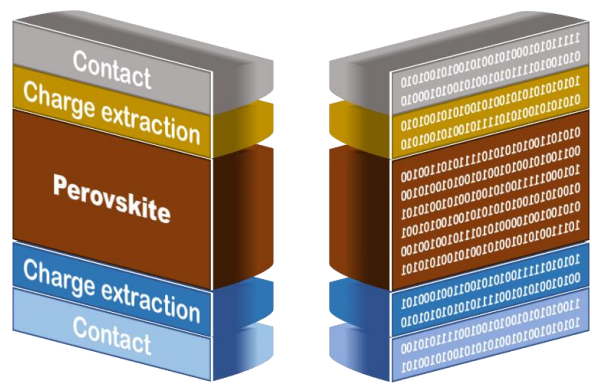

MAIN TEXT

Device modeling and simulation are powerful tools that have furthered the development of many electronic technologies.[1-5] They provide significant insights that can be used to complement experimental results, but are not always directly available from empirical studies. Experimentally, the implications of changing a single parameter in the device structure or fabrication procedure are often complex and even unpredictable, such that assigning the changes in device performance to a single source is challenging and sometimes impossible, a limitation that simulations do not possess. It is this interplay between material properties and device parameters in determining the device performance that makes modeling and simulation of devices so attractive to material scientists, physicists, and engineers.

In the field of emerging photovoltaic technologies based on organic materials, device modeling has been extensively used to not only develop a deeper understanding of the physical principles governing the device function,[6-8] but also as a tool for optimizing the device structure in order to enhance its performance.[9-10] Similarly, albeit to a lesser degree, device simulations have also been applied to colloidal quantum dot photovoltaics.[11]

Since the advent of perovskite photovoltaics, significant experimental efforts have been devoted to enhancing the device power conversion efficiency (PCE), reaching remarkable improvements in just a decade.[12] This continuous improvement in PCE is a result of optimization in perovskite film composition,[13-14] deposition procedures,[15-16] device architecture,[17] charge extraction [18-20] and charge blocking layers.[21-22] Accompanying this remarkable experimental success are somewhat limited efforts in device modeling and simulation. The vast majority of these efforts have been dedicated to developing an understanding of the hysteresis effects[23-26] and the transient behavior observed in perovskite-based solar cells.[27-28] The key difference between these simulations and those developed earlier for organic or quantum dot photovoltaic devices is the inclusion of ions as species that can move upon the application of an electric field. This means that an additional set of drift-diffusion equations for cations and anions has to be added and solved numerically. Similarly, the Poisson equation also needs to be modified to account for the ions' charge. This additional complexity might deter researchers from attempting to develop a comprehensive device model for perovskite solar cells. In the following, we will discuss the role of ions within the device simulation and identify in which cases they must, or should, be included and when they can be left out from the simulations entirely.

Another aspect of device structure that has been addressed by simulation is the effect of the perovskite's energetic alignment with the charge extracting layers of the device.[29-34] These energetic barriers influence the boundary conditions for electrons and holes in the simulation and are of interest experimentally due to the wealth of charge extraction materials used in devices. We will focus on energetic misalignment as an example, which allows us to explore a range of simulation parameters and identify key factors that must be taken into account.

The effects of the lateral[35] and vertical[36] distribution of grains and grain boundaries has also been simulated, however a far deeper investigation of their role is needed, especially in light of recent experimental observations that 'grains', as observed by scanning electron microscopy, do not always correctly correspond to the crystallographic grains within the perovskite film. Imaging of crystallographic perovskite grains by electron backscatter diffraction suggests that previous reports might have overestimated the grain size of perovskite films.[37] This and other conflicting reports concerning the influence of grain sizes and grain boundaries on the properties and performance of perovskite solar cells underline that the their role remains under debate.[38] Finally, minor effort has been devoted to simulation of device stability[39-41] – another challenge in which device modeling can be a major source of insight.

In this perspective, we would like to introduce the basics of device modeling and highlight the key parameters that must be taken into account for such simulations to be particularly useful to the perovskite community. Our goal is to motivate the application of device modelling as a powerful tool to complement experimental efforts to understand the device physics of perovskite solar cells and further enhance their performance.

*The basic principles of device modelling*

To model perovskite-based photovoltaic devices, the following set of drift-diffusion equations has to be solved numerically:

$$\begin{cases} \frac{\partial}{\partial t} n_e(z,t) = \frac{\partial}{\partial z}\left[ D_e(n_e)\frac{\partial}{\partial z} n_e + \mu_e n_e E \right] - B n_e n_h + \alpha G(z,t) \\ \frac{\partial}{\partial t} n_h(z,t) = \frac{\partial}{\partial z}\left[ D_h(n_h)\frac{\partial}{\partial z} n_h - \mu_h n_h E \right] - B n_e n_h + \alpha G(z,t) \\ \frac{\partial}{\partial t} n_c(z,t) = \frac{\partial}{\partial z}\left[ D_c(n_c)\frac{\partial}{\partial z} n_c + \mu_c n_c E \right] \\ \frac{\partial}{\partial t} n_a(z,t) = \frac{\partial}{\partial z}\left[ D_a(n_a)\frac{\partial}{\partial z} n_a - \mu_a n_a E \right] \end{cases} \quad (1)$$

where:

$n_e$, $n_h$, $n_c$, $n_a$ are the densities (cm$^{-3}$) of electrons, holes, cations and anions, respectively.

D, µ, and E are the diffusivity, mobility and electric field.

B is the bimolecular recombination coefficient αG is the generation rate, where α accounts for not all photons converting into free charges. One may add traps and trap assisted recombination either in the bulk or at the interfaces; these will be discussed in more detail later. The numerical scheme to include the generalized Einstein relation can be found in ref 42.

The above drift-diffusion equations are coupled to the Poisson equation, which accounts also for the ions' charge:

$$\frac{d}{dz}E = \frac{q}{\varepsilon(z)\varepsilon_0}\left(n_h(z) - n_h(z) + n_a(z) - n_c(z)\right) \qquad (2)$$

The boundary conditions for electrons and holes at the contacts are:

$$\begin{cases} n_e(0) = N_{e0}\exp\left(-\frac{\Delta_e}{kT}\right) & ; n_e(L) = N_{e0}\exp\left(-\frac{\Delta_h}{kT}\right) \\ n_h(0) = N_{h0}\exp\left(-\frac{E_{g1}-\Delta_e}{kT}\right) & ; n_h(L) = N_{h0}\exp\left(-\frac{E_{g3}-\Delta_h}{kT}\right) \end{cases} \qquad (3)$$

Here:

$N_{e0}$ and $N_{h0}$ are the effective density of states for electrons and holes, respectively.

$\Delta_e$ and $\Delta_h$ are the energy separation between the LUMO level and the contacts' Fermi level at the cathode and anode, respectively.

$E_{g1}$ and $E_{g3}$ are the energy gaps of the semiconductor layers next to the cathode and anode, respectively. For the ions, the boundary conditions at the perovskite interface are taken to be blocking. When implementing the above equations one needs to account for discontinuities and in the present context one needs to also pay attention to the density of states[43] and the dielectric constant.[44-45]

Beyond the computational challenge of solving these equations, it is clear that the choice of certain parameters will greatly influence the outcome of the simulation. In the following, we will discuss some of these influences and motivate the choice of certain parameters for perovskite device simulation. To consistently demonstrate the influence of certain parameters, we will perform all the simulations shown in this prospective, rather than attempt to compare previous simulation results performed by other groups, which use different sets of parameters and are thus not directly comparable. The parameters used here are listed in Table 1.

**Table 1**. Material properties used in the simulations (unless stated otherwise)

|  | Active layer | Blocking layers |
|---|---|---|
| Electron/hole mobility ($cm^2v^{-1}s^{-1}$) | 2 | 0.01 |
| Bimolecular recombination ($cm^3s^{-1}$) | $10^{-10}$ | Langevin |
| $N_C$ ($cm^{-3}$) | $7 \times 10^{18}$ | $10^{21}$ |
| $N_V$ ($cm^{-3}$) | $2.5 \times 10^{18}$ | $10^{21}$ |
| $\varepsilon$ | 50 | 3 |
| Anion diffusivity | ----- | ----- |
| Cation diffusivity ($cm^2s^{-1}$) | $10^{-12}$ | ----- |
| Ion density ($cm^{-3}$) | $10^{18}$ | 0 |
| Thickness (nm) | 250 | 50 |

*Role of ions - in or out?*

The mixed conductor nature of perovskite materials suggests that both ions and charges are moving upon the application of an electric field.[46] The densities and dynamics of ions are different to those of charges, suggesting that their migration and electrostatic interaction will influence both the temporal evolution and the steady-state current in the device. It is also noteworthy that theoretical and experimental measurements suggest that the stark difference in diffusion coefficients of the cations and anions[47-49] in perovskite materials effectively means that only one species is moving, which at first glance should introduce a significant asymmetry into the device.

While it is clear that modeling of transient phenomena must include the movement of ionic species, numerical simulation of solar cell current-voltage characteristics, especially in the absence of hysteresis, can in principle be performed either with or without the inclusion of ions. It is known that even in the absence of hysteresis the ions are still moving and affecting the charge distribution across the device.[50] Hence, it is not clear if simulating experimental data using a model that excludes ions is capable of reproducing the correct physical picture (e.g. the type of recombination processes and/or their spatial location/distribution). We chose to answer this question while examining the relationship between the built-in and open-circuit voltages in perovskite solar cells. This relation has been under debate and despite extensive numerical studies of perovskites cells,[29,51-52] this specific issue hasn't been addressed in detail.

Generally, the built-in voltage ($V_{BI}$) of a solar cell will depend on the energetic difference between the Fermi level positions of the cathode and the anode. In the presence of charge extraction layers, these levels are typically pinned within the gap of the extraction layers: a few 100 meV above the hole transport level or below the electron transport level for the anode and cathode, respectively. This means that for the ideal case of perfect energetic alignment between the charge transport levels of the extraction layers to those of the perovskite active layer, the built-in voltage is still lower than the optical gap of the active layer. Additionally, in reality, there are often energetic offsets between the charge transport levels of the extraction layers and those of the perovskite, decreasing the built-in voltage even further. Conflicting results regarding the importance of these energetic offsets have been reported in literature. Several studies showed that by increasing the ionization potential of the hole-transporting layer, a higher open-circuit voltage ($V_{OC}$) can be attained due to a more optimal alignment with the perovskite valence band.[53-55] On the other hand, some publications show very little correlation between these two values.[56-57] Similar inconsistencies exist in relation to the energetic offset with the electron extraction layer.[58-59]

To investigate numerically the impact of such energetic offsets and verify if excluding the ions still allows to correctly reproduce the internal mechanisms, we perform two sets of simulations. We use identical sets of device and material parameters and run the simulations with and without the inclusion of ions. For the case where ions are included, we simulate a hysteresis measurement at a scan speed of 50 mV/s, which when normalized to the ion diffusivity, results in $5\times10^{13}$ mVcm$^{-2}$. When one performs a hysteresis loop measurement one may miss capturing the hysteresis by either scanning very slow such that the ions closely follow the voltage change or very fast such that the ions do not move during the forward and reverse voltage sweep. To avoid the latter issue, we separate the loop into two scans. First, the cell is kept at low bias long enough to allow the ions reach steady state and then a forward scan is performed. For the reverse direction the device is held at a high bias voltage, long enough for the ions to reach steady state, followed by a backward voltage sweep. Using this methodology implies that faster scan rates are more likely to lead to the observation of hysteresis. For the current simulations, the stabilization bias for the forward and reverse scans are -0.15 V and 1.25 V, respectively. The energy level diagram for both scenarios, of with and without ions, is shown in Figure 1a. For simplicity, we use a 1.6 eV bandgap for the perovskite active layer as well as anode and cathode fermi levels, which are pinned 0.2 eV below/above the electron/hole transport levels. Such a configuration means that in the case of perfect energetic alignment between the extraction layers and the perovskite active layer, the maximum built-in voltage is 1.2 V. We now introduce an energetic offset, $\Delta$, between the valence band of the perovskite layer and the highest occupied molecular orbital (HOMO) of the hole extracting layer, while maintaining perfect alignment with the electron extracting layer. Correspondingly, the built-in potential in the device is lowered by $\Delta$, making it 1.2-$\Delta$ eV. The relationship between the simulated open-circuit voltage and this built-in voltage, with and without including ions, is shown in Figure 1b and leads to several fascinating observations. First, for any energetic offset, $V_{OC}$ always surpasses $V_{BI}$. Interestingly, in the case of optimal alignment (i.e. $\Delta$=0 eV), a $V_{OC}$ of 1.23 V is predicted, which is only 0.03 V higher than the built-in potential. However, in the case of misalignment the $V_{OC}$ is ~0.35 V higher than the built-in potential, which shows the remarkable recovery of $V_{OC}$ in perovskite solar cells even in the case of large energetic offsets. Secondly, very strikingly, simulations performed without the contribution of ions show near identical results. In fact, the J-V curves of both the forward and the reverse scans overlap not only with each other, but also with the J-V obtained when no ions are included.

To understand the origin of the $V_{OC}$ recovery, we turn our attention to the charge density distributions at $V \approx V_{OC}$ for the case of $\Delta = 0.3$ eV, shown in Figure 1c and d. Figure 1c shows the charge distribution for forward (dashed line) and reverse (solid line) scans when ions are included, while Figure 1d is for ions being excluded. While the distributions for these three scenarios are markedly different, they do share a common feature – the formation of a dipole at the perovskite/HTL interface. This interfacial dipole is a result of an excitation generated high density of electrons and holes on the perovskite and HTL sides, respectively. In the results presented above, the inclusion of ions has only a minimal effect: while they help create the dipole, it is primarily electronic in nature. Such a self-induced dipole is made possible by the very low losses in perovskite semiconductors, and as such is a unique feature of perovskites among other emerging semiconductors. The resulting energy level diagrams for the case where ions are included (reverse scan) and the no-ions case are shown in Figure 1e and 1f. This dipole reduces the effective energy offset and thus significantly enhances the effective built-in potential of the device, explaining the high values of the simulated $V_{OC}$s even for large energetic offsets.

Upon illumination an **interfacial, primarily electronic, dipole** is formed in perovskite solar cells with **non-ideal energetic alignment**. This self-induced dipole strongly increases the built-in potential and device open-circuit voltage, **suppressing the losses** introduced by the misalignment and is a **unique feature** of low recombination perovskite semiconductors.

The fact that the two simulations (with and without ions) employ the same physical parameters and result in the same J-V characteristics supports the idea that neglecting the ions still allows to capture the correct physics (for devices with no hysteresis). This, however, could be associated with the relatively simple physical picture used in this example and one could argue that accounting for additional mechanisms would invalidate such a conclusion. Such an argument is potentially supported by the observation that unlike the similar densities of electrons and holes at the perovskite/HTL interface, the distribution of charges in the bulk of the perovskite active layer is very different depending on scan direction or whether ions are included or not. For the reverse scan the ions were stabilized at 1.25V which is close to $V_{OC}$ and hence they almost completely suppress the electric field in the bulk, generating an effective 'flat band' state. This results in the density of both charge carriers being nearly constant throughout the bulk of the perovskite layer.

In the forward scan, the scan speed does not allow the ions to fully screen the electric fields at $V_{OC}$ and naturally, without ions there is no screening at all. In the absence of ion-screening, the distributions of electrons and holes vary greatly in the bulk of the active layer. Similarly, at the perovskite/ETL interface, up to one order of magnitude difference in the density of charges occurs between the different scenarios. The stark difference between the charge carrier distribution curves (solid lines in Figure 1c and 1d) might challenge the notion that excluding the ions is justified regardless of the existing internal mechanisms or their spatial distribution. We argue, however, that this concern is alleviated by comparing the charge carrier density distributions for the forward and reverse scans in the case ions are included. As Figure 1c shows, these charge density distributions are similarly different, and yet they did not result in different current densities (no hysteresis), confirming that the processes taking place in the device are not sensitive to the charge density distribution. Thus, it is justified to use a model that does not account for ions to simulate the internal mechanisms of a device that does not exhibit hysteresis despite the presence of ions.

To sum up the above observations: if one scans the J-V at a normalized speed of at least $5 \times 10^{13}$ mVcm$^{-2}$ (or 50 mV/s for ion diffusivity of $10^{-12}$ cm$^2$/s), the ions are mobile enough to significantly alter the charge density distribution between the forward and reverse scans (note the unique scan procedure established above). If this is not accompanied by (noticeable) hysteresis in the J-V curve, then modeling the device with or without the inclusion of ions yields (almost) identical results. This conclusion does not require that there will not be hysteresis at any scan speed, and it may indeed appear at scan speeds of volts per second. However, the scan speed should not be too slow, which is why we provided a reference value of 50 mV/s. Consequently, modeling of device characteristics, which do not exhibit a clear signature of ions (e.g. hysteresis) can be done without the inclusion of ions with near identical results. Moreover, the fact that only one species of ions is moving does not affect the device symmetry – an accumulation of ions of one type at a certain interface is always accompanied by the formation of a deficiency at the other interface. This accumulation and deficiency will be of opposite charge signs, meaning that movement of neither anions nor cations nor both will introduce asymmetry into the device.

The results of the numerical simulations show that small energetic offsets (Δ ≤ 0.3 eV) introduce nearly no losses in $V_{OC}$ since the self-induced dipole can effectively compensate for this non-ideality. Larger offsets negatively impact the $V_{OC}$, but their effect is mitigated by the self-induced dipole by approximately 0.35 V. This observation is of critical importance for real life devices, since it significantly broadens the range of extraction layers that can be employed in perovskite photovoltaics, at least with respect to the requirements of their energetic alignment to the perovskite active layer.

> Simulation of J-V characteristics that show **no signs of hysteresis** can be performed **without including ions** in the device model and still reproduce the same material and device parameters.

When considering these results in light of the reported inconsistencies in literature regarding the role of energetic alignment in determining the device open-circuit voltage, one has to remember that experimentally changing the extraction layer varies not only the energetic alignment with the perovskite layer, but also many other factors. For example, it has been shown that the charge carrier mobility of the extraction layer,[60-61] the interfacial recombination losses,[62] and the energetic disorder within the extraction layer[63] can all influence the photovoltaic performance. Moreover, comparing charge extraction layers on top of which the perovskite layer is deposited is further complicated by changes in the microstructure and defect density of the perovskite depending on the surface energy, roughness, and other properties of the underlying extraction layer.[19,64] In fact, in the simulations presented herein, we chose a mobility of $10^{-2}$ $cm^2V^{-1}s^{-1}$, such that under no circumstances would transport within the EEL and HEL limit charge extraction, thus ensuring that the only barrier imposed on the system is introduced by the energetic misalignment. We note that we chose to use undoped charge blocking layers, which become increasingly common due to the potential benefit to device stability.[26,65]

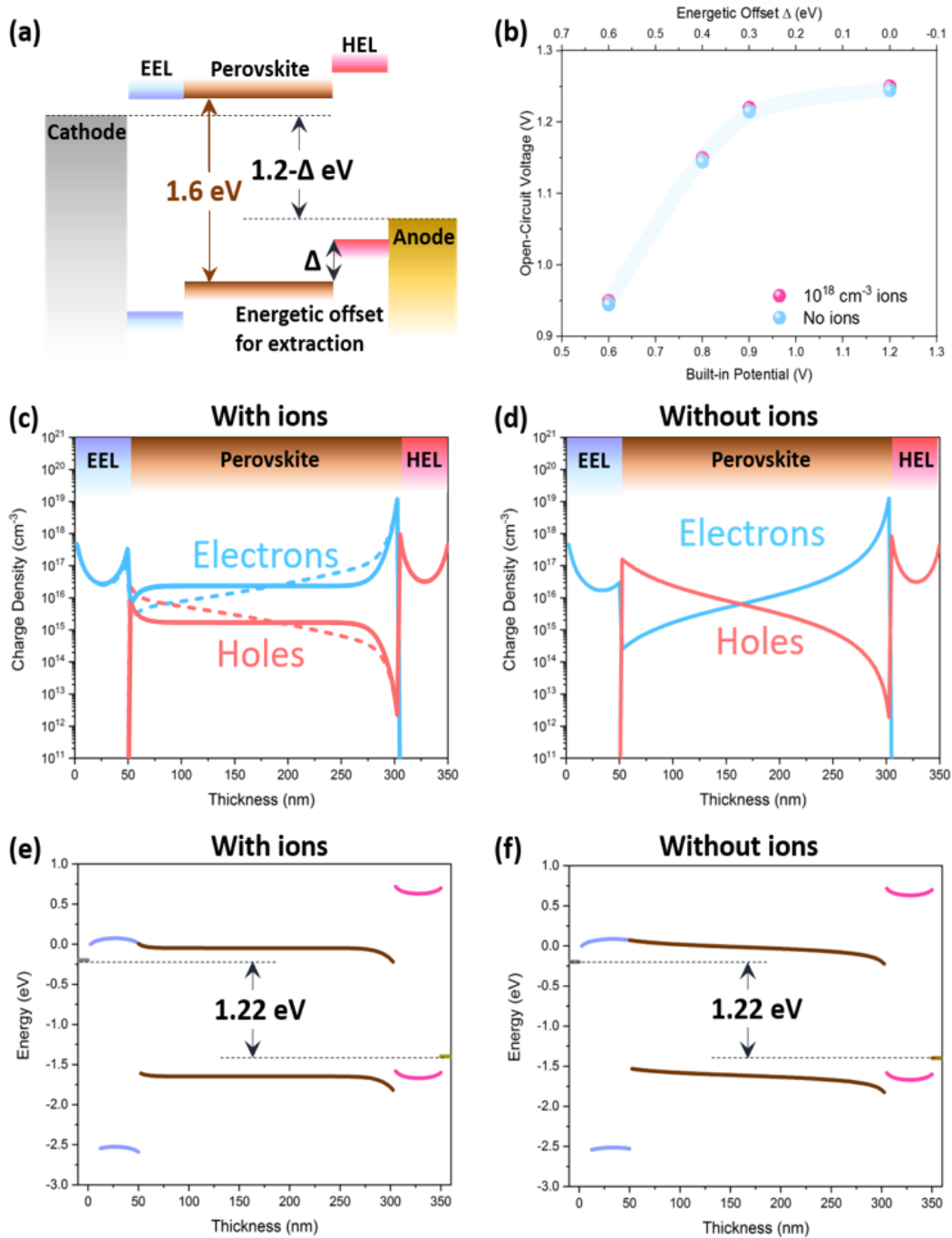

**Figure 1**: (a) Energy level diagram used for modeling of the effect of energetic offset at the hole-extracting layer (HEL). (b) Simulated open-circuit voltage as a function of the built-in potential. Charge density distribution at V ≈ $V_{OC}$ for Δ = 0.3 eV (c) with and (d) without the inclusion of ions. In (c) dashed and solid lines are for forward and revers scans (50 mV/s), respectively. Energy level diagrams at V≈$V_{OC}$ for Δ = 0.3 eV (e) with ions and in reverse scan and (f) without the contribution of ions.

*Role of the dielectric constant*

Regardless of whether one chooses to include the ions or not, the dielectric constant of the perovskite active layer is a key material parameter that needs to be known in order to solve the Poisson equation (see equation 2) and has been shown to affect simulations of device performance.[23] Despite a decade of research, little agreement exists regarding this value for a range of perovskite materials. Theoretical calculations and various experimental measurements have resulted in values that vary across an order of magnitude both among different perovskite compositions and even for the same composition.[66]

To exemplify the influence of the dielectric constant on the simulated device performance, we continue the example presented above (Figure 1), in which we examine the effect of an energetic offset on photovoltaics characteristics of a solar cell. Following literature reports on the possible values of the dielectric constant, we chose either $\varepsilon = 6$ or $\varepsilon = 50$, with all other model parameters kept unchanged (Figure 2). For the ideal case of perfect alignment between the active layer's and extraction layers' transport levels, i.e. $\Delta = 0$ eV, a change in the dielectric constant has no effect on the J-V characteristics. With no losses in $V_{BI}$ to overcome, only the balance between generation and recombination dictates the photovoltaic performance, which are not influenced by the dielectric constant (the potential effect on exciton dissociation through its binding is not addressed). However, when an energetic offset is present, the value of the dielectric constant plays a major role: low $\varepsilon$ leads to significantly better photovoltaic performance with particularly high $V_{OC}$s. For example, for the highest energetic offset of $\Delta = 0.6$ eV, a $V_{OC}$ of 1.17 V is obtained, which corresponds to a remarkable recovery by 0.57 V. The reason for this is evident in the Poisson equation itself – for low values of $\varepsilon$ the effect of charges on the energy bands is significantly enhanced.

These results have two key implications: first, they show that the choice of ε for device modeling studies is of crucial importance, but far more importantly, it provides a critical insight into the device physics of perovskite solar cells, demonstrating the astonishing and surprising advantage of engineering perovskite materials with lower dielectric coefficients. It is generally considered that high dielectric coefficient values are desirable in order to lower the exciton binding energy in the photoactive material, however our results indicate that in the case of a non-ideal device structure, this can also lower the photovoltaic performance. With the ability to modify the dielectric properties of perovskites by varying their composition, it should be possible to balance the two effects and develop perovskites with sufficiently low exciton binding energy (i.e. comparable to kT), but also sufficiently low dielectric constants such that high photovoltaic performance can be obtained for a broad range of non-ideal device architectures.

> The exact value of the perovskite dielectric constant is **critically important** for device simulations. A low dielectric constant enhances the effect of the self-induced dipole, resulting in better photovoltaic performance for a broader range of energetic misalignments.

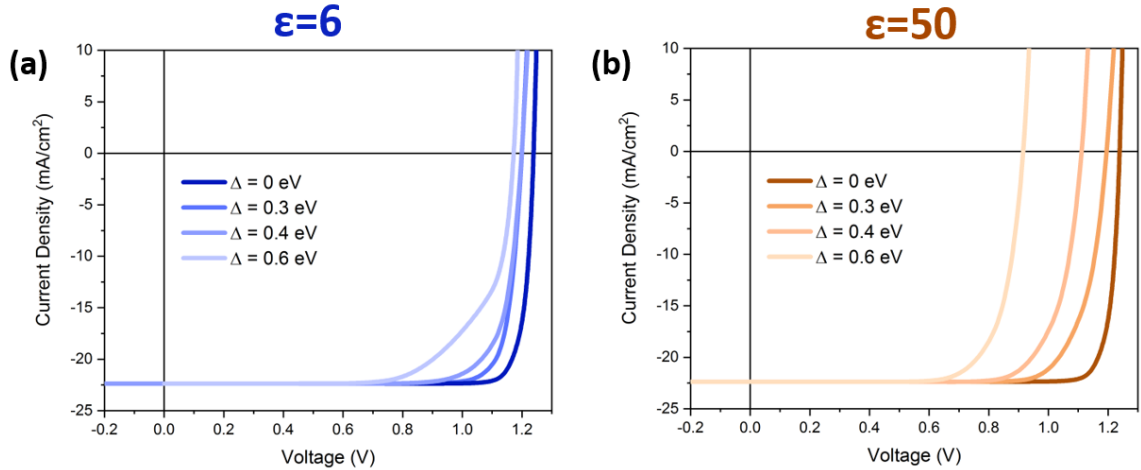

**Figure 2**: J-V characteristics for varying HEL offsets (Δ = 0, 0.3, 0.4, 0.6 eV) for dielectric coefficient of (a) 6 and (b) 50.

*Role of the density of states*

While often neglected from discussion, the density of states plays an important role in determining the photovoltaic performance of materials. The effective density of states at the band edges, $N_{e0}$ and $N_{h0}$ in equation set (3), will dictate how close the Fermi level will approach the band when populated with a certain density of charges. Correspondingly, and perhaps counter intuitively, a lower effective density of states results in a higher $V_{OC}$ (Figure 3a,b). This important result was reported by Zhou et al,[67] but has gone largely unnoticed by the perovskite community. It is noteworthy that experimental studies indeed suggest that the density of states at the band edges of commonly used perovskites are rather low.[68] As discussed in ref 67, the effective density of states at the band edge represents the entire density of states distribution function as a single parameter that allows to reproduce the charge density dependence on the Fermi level position using the Boltzmann approximation (where such approximation is valid). The specific values reported in 67 make use of the fact that for parabolic bands one can derive a relation between the charge effective mass and the effective density of states.

The population of the electronic states is also important. Most commonly, Boltzmann statistics are used to determine this population, but similarly to organic semiconductors,[69] one has to consider whether Boltzmann statistics is valid or whether the full Fermi-Dirac (FD) statistics should be applied. FD statistics are often implemented with the aid of the Generalized Einstein Relation (GER), which is given by:

$$\frac{D}{\mu} = \frac{p}{q\frac{\partial p}{\partial \eta}} \qquad (4)$$

Where p is the particle concentration and η is the chemical potential.

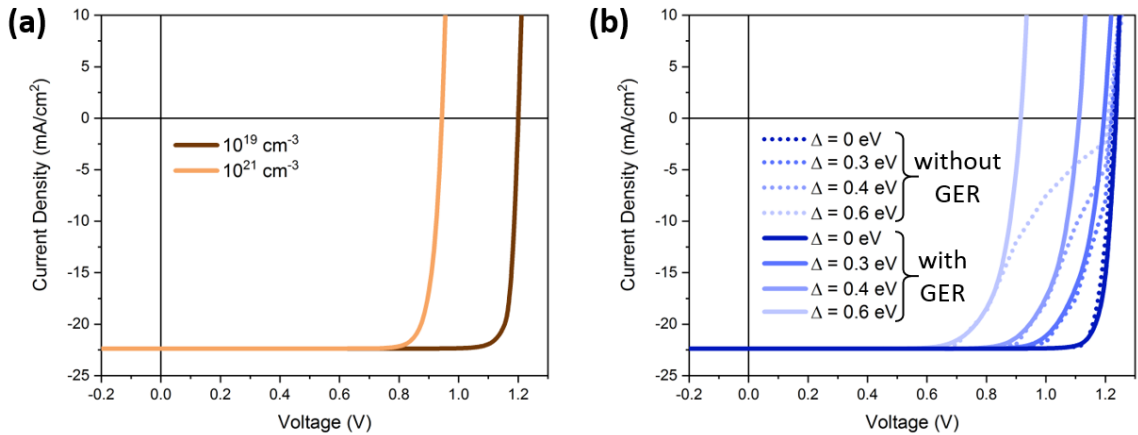

**Figure 3**: (a) Simulated J-V characteristics for $N_{e0}$ of $10^{19}$ and $10^{21}$ cm$^{-3}$ and Δ = 0 eV. (b) Simulated J-V characteristics with and without GER (in this example only, the simulations were run in the slow scan limit where ions are allowed to reach the steady state at each bias).

Continuing with the example given above, we explore how neglecting the GER (i.e. using Boltzmann instead of Fermi-Dirac) influences the photovoltaic characteristics of simulated perovskite devices. Figure 3b shows that neglecting the GER, which removes the effective density of states limitation on the charge density, results in higher interfacial charge carrier densities and larger self-induced dipoles, thus fully recovering the $V_{OC}$. This of course is a highly unrealistic scenario, indicating that FD statistics, implemented using GER, cannot be ignored when considering density of state population in perovskite semiconductors for non-ideal device structures.

> The full **Fermi-Dirac statistics**, or Generalized Einstein Relation, **must be considered** when simulating perovskite based solar cells.

The observation that the GER or FD statistics must be employed for simulations of perovskite solar cells is related to their low effective density of states at the band edge, mentioned above. At $V_{OC}$, the formation of the self-induced dipole at the perovskite/HEL interface results in a high density of electrons (Figure 1c, d), that in combination with the low effective density of states places the quasi Fermi level very close to the band edge. This effectively degenerate state of the perovskite semiconductor at the interface renders Boltzmann statistics no longer valid, making it necessary to include GER (i.e. FD statistics) into the simulations.

*Role of bulk and interface recombination*

While it is generally agreed upon that recombination losses in perovskite materials are far lower than in other emerging semiconductors, the effects of bulk and interfacial recombination on the performance of perovskite photovoltaics are of great interest to the scientific community.[70] Modeling is an excellent tool to investigate the effects of recombination,[23,28] since both its nature and physical location in the device can be chosen at will. We follow up on our example from Figure 1, but this time increase the bulk recombination by two orders of magnitude (Figure 4a). Several critical observations can be made. As expected, the $V_{OC}$ of the device with an ideal energetic alignment ($\Delta = 0$ eV) is significantly decreased compared to the prior case of low bulk recombination, and is now comparable to what previously was attained for $\Delta = 0.4$ eV. Interestingly, this means that the introduction of an energetic offset of up to $\Delta = 0.4$ eV no longer has an effect on the $V_{OC}$, but has a tremendous negative effect on the fill factor of the device, manifesting itself as an 'S-shape' in the J-V characteristics. Moreover, the presence of high bulk recombination results in the appearance of hysteresis, which is particularly enhanced in non-ideally aligned devices.

Figure 4b shows the results obtained when the recombination is restricted to the interface between the perovskite and the two extraction layers, with the bimolecular recombination coefficient being increased to result in the same $V_{OC}$ for the $\Delta = 0$ eV case as in Figure 4a. Namely, we only probe the effect of the spatial distribution of the losses, without altering the effective loss amount or the loss mechanism. Due to the relatively high mobility of the charge carriers, the spatial distribution of the losses has almost no effect on the charge density distribution across the device, resulting in very similar J-V curves for both the bulk and interfacial recombination cases. While the spatial distribution may slightly alter the shape of the J-V curves, it does not determine the presence of hysteresis.[71] We note that interfacial recombination would have a significant effect on device hysteresis in cases where it serves as an additional loss mechanism or is associated with a high trap density.[23] As has been reported for some memory devices,[72] the filling of the traps may form a space charge that would diminish the free charge density in their vicinity, and thus limit or delay the current flow. It is also noteworthy that while the resulting $V_{OC}$s are generally smaller than in the case of low recombination (Figure 1), they still surpass the built-in potential, suggesting that the self-induced dipole still compensates for some of the losses. In this case, however, it is expected that the contribution of ions to the dipole becomes more important, since the high recombination rate lowers the density of charges developed at that interface. This is examplified by simulations for Δ = 0.6 eV which exclude the ions (Figre 4a, dotted line). As expected, the Voc is indeed lower than when ions are taken into account, confirming their contribution to the interfacial dipole. However, this contribution is relatively small, suggesting that the dipole remains primarily electronic in nature even in the case of high recombination losses. It is particularly important to note that excluding the ions does not result in J-V curves that resemble the average between forward and reverse scans. This epitomizes that attempting to reconstruct the J-V characteristics of a device with hysteresis using a simulation that excludes ions will inevitably lead to inaccurate interpratation of the internal device.

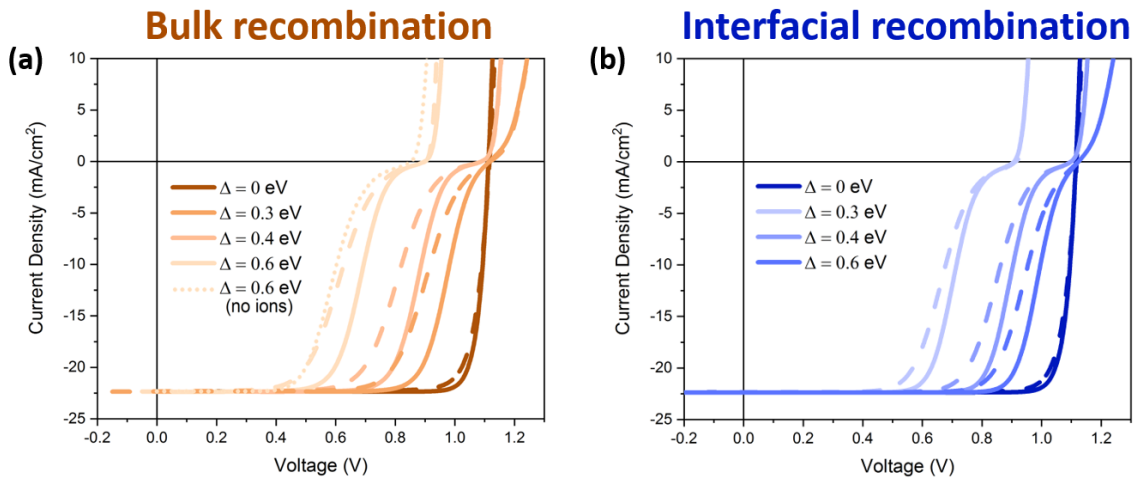

**Figure 4**: J-V characteristics for varying HEL offsets (Δ = 0, 0.3, 0.4, 0.6 eV) for increased (a) bulk and (b) interfacial recombination. Dashed lines represent the forward scan. Scan speed was 50 mV/s and the stabilized voltages were -0.15 V and 1.25 V for the forward and reverse scan, respectively. The dotted line in (a) was computed when neglecting the presence of ions for the case Δ = 0.6 eV.

> **Increased recombination losses**, either bulk or interfacial, result in **S-shaped current-voltage** characteristics, a loss in $V_{OC}$ and FF, and the **appearance of hysteresis**. To reproduce the internal mechanisms, devices in which large recombination losses are present must be simulated with the contribution of ions taken into account.

*Simulating interfacial engineering for $V_{OC}$ enhancement*

With the formation of a self-induced dipole largely suppressing the limitations imposed by the non-ideal energetics and low built-in potential, one could consider that the introduction of additional dipoles in the solar cell by interfacial engineering might be of benefit for $V_{OC}$ enhancement. Indeed, we and others have demonstrated that introducing dipoles into device architectures can significantly enhance its $V_{OC}$.[21,73-75] Device simulation incorporating external dipoles into the device structure showed that increasing the dipole strength up to 0.4 eV results in a linear increase in the $V_{OC}$ which is proportional to the dipole strength. Experimental studies show that such dipoles can result in an increase in $V_{OC}$ of up to 120-130 mV without any effect on the $J_{SC}$ or FF of the device, resulting in an overall enhancement in the PCE on the order of ~10%.

*Future Directions of Perovskite Solar Cell Modelling*

In this perspective we describe the consequences of selecting certain aspects and parameters for perovskite device simulation, such as inclusion/exclusion of ions or the type of occupation statistics. We would like to emphasize that it is not categorically "wrong" to use Boltzmann instead of Fermi-Dirac statistics just as it is not "wrong" to not include ions. We demonstrated that under certain scenarios such a choice might be warranted. For example, we provide the presence of hysteresis as the criterion for when neglecting the ions may lead to an erroneous physical picture. Opting to use Boltzmann statistics in cases where the simulation predicts charge densities close to, or above, the effective density of states would also lead to inaccurate results. These issues, highlighted in this work, signify the early stage of perovskite device modeling where efficient codes that can handle mixed electronic-ionic conduction are not widely available. On the other hand, there are many other potential mechanisms that have not been fully addressed yet,[76-77] partly because they are not understood well enough to be included. For example, how to account for the coupling between the ion accumulation/deficiency and electronic traps[28,39] - would these be efficient recombination traps?[78] Should light activation of ionic transport[79-80] be incorporated into the device model, namely would the light excitation profile within the device affect the ionic distribution in a manner reminiscent of the effects of thermal gradients on electronic transport? How should one describe the grain boundaries? Can they be considered as only affecting the electrical properties of the grain, or does one have to use 2D modeling so that the boundaries are treated separately? For stability studies,[81] current models are clearly missing ionic transport through the blocking layers[82] and any potential electrochemistry at the contacts.[83-84] Furthermore, it has been suggested that perovskite materials may exhibit self-healing processes[85-86] - how would such processes affect the ionic and electronic landscape? Is a certain amount of disorder needed due to the interplay between degradation and self-healing?

All of the above questions, and others that will emerge as more information becomes available about the physics and chemistry of perovskite materials, are important open questions in the field. Some of these questions will only be answered through the use of device models to help separate the effects of various processes and suggest which description of a given physical mechanism is more accurate in terms of allowing the device model to reproduce experimental device data. Future studies of the missing mechanisms, and their translation into a set of equations that can be incorporated into a device model, would make a significant difference in the future progress of the field of perovskite photovoltaics.

Efficient, flexible and open source code models will enable not only the development of fundamental understanding of device physics of perovskite solar cells, but would also serve as device design guidelines to enhance the performance and stability of devices, reducing the current 'trial and error' experimental approach for advancing the field towards industrial application.

AUTHOR INFORMATION

**Notes**

The authors declare no competing financial interest.

ACKNOWLEDGMENT

This project has also received funding from the European Research Council (ERC) under the European Union's Horizon 2020 research and innovation program (ERC Grant Agreement n° 714067, ENERGYMAPS). N.T. acknowledges support by the Israel Science Foundation (grant no. 488/16), the Adelis Foundation for renewable energy research within the framework of the Grand Technion Energy Program (GTEP), and the Technion Ollendorff Minerva Center.REFERENCES

[1] Sze, S. M., Physics of Semiconductor Devices; Wiley: New York, 1981.

[2] Street, R. A., Technology and Applications of Amorphous Silicon; Springer-Verlag: Berlin, 1999.

[3] Selberherr, S., Analysis and Simulation of Semiconductor Devices; Springer-Verlag: Vien, 1984.

[4] Snowden, C. M., Introduction to Semiconductor Device Modelling; World Scientific, 1998.

[5] Kao, K. C.; Hwang, W., Electrical Transport in Solids; Pergamon press: New York, 1981; Vol. 14.

[6] Tessler, N.; Rappaport, N., Excitation Density Dependence of Photocurrent Efficiency in Low Mobility Semiconductors. *J. Appl. Phys.* **2004**, 96, 1083-1087.

[7] Koster, L. J. A.; Smits, E. C. P.; Mihailetchi, V. D.; Blom, P. W. M., Device Model for the Operation of Polymer/Fullerene Bulk Heterojunction Solar Cells. Phys. Rev. B 2005, 72, 085205.

[8] Kirchartz, T.; Nelson, J., Device Modelling of Organic Bulk Heterojunction Solar Cells. *Top. Curr. Chem.* **2013**, 352, 279.

[9] Tessler, N., Adding 0.2 V to the Open Circuit Voltage of Organic Solar Cells by Enhancing the Built-in Potential. J. Appl. Phys. 2015, 118, 215501.26

[10] Tan, J.-K.; Png, R.-Q.; Zhao, C.; Ho, P. K. H., Ohmic Transition at Contacts Key to Maximizing Fill Factor and Performance of Organic Solar Cells. *Nat. Commun.* **2018**, 9, 3269.

[11] Solomeshch, O.; Kigel, A.; Saschiuk, A.; Medvedev, V.; Aharoni, A.; Razin, A.; Eichen, Y.; Banin, U.; Lifshitz, E.; Tessler, N., Optoelectronic Properties of Polymer-Nanocrystal Composites Active at near-Infrared Wavelengths. *J. Appl. Phys.* **2005**, 98, 074310.

[12] NREL, Best research-cell efficiencies. Accessed on 05/12/2019.

[13] Saliba, M.; Matsui, T.; Seo, J.-Y.; Domanski, K.; Correa-Baena, J.-P.; Khaja Nazeeruddin, M.; Zakeeruddin, S. M.; Tress, W. R.; Abate, A.; Hagfeldt, A.; et al. Cesium-containing triple cation perovskite solar cells: improved stability, reproducibility and high efficiency. *Energy Environ. Sci.* **2016**, 9, 1989-1997.

[14] Saliba, M.; Matsui, T.; Domanski, K.; Seo, J.-Y.; Ummadisingu, A.; Zakeeruddin, S. M.; Correa-Baena, J.-P.; Tress, W. R.; Abate, A.; Hagfeldt, A.; et al. Incorporation of rubidium cations into perovskite solar cells improves photovoltaic performance. *Science* **2016,** 354, 206-209.

[15] Guo, F.; Qiu, S.; Hu, J.; Wang, H.; Cai, B.; Li, J.; Yuan, X.; Liu, X.; Forberich, K.; Brabec, C. J.; et al. A Generalized Crystallization Protocol for Scalable Deposition of High-Quality Perovskite Thin Films for Photovoltaic Applications. *Sci. Adv.* **2019**, 6, 1901067.

[16] Eggers, H.; Schackmar, F.; Abzieher, T.; Sun, Q.; Lemmer, U.; Vaynzof, Y.; Richards, B. S.; Hernandez-Sosa, G.; Paetzold, U. W. Inkjet-Printed Micrometer-Thick Perovskite Solar Cells with Large Columnar Grains. *Adv. Energy Mater.* **2019**, Early view.

[17] Liu, T.; Chen, K.; Hu, Q.; Zhu, R.; Gong, Q. Inverted Perovskite Solar Cells: Progresses and Perspectives, *Adv. Energy Mater.* **2016**, 6, 1600457.

[18] Le Corre, V. M.; Stolterfoht, M.; Perdigon Toro, L.; Feuerstein, M.; Wolff, C.; Gil-Escrig, L.; Bolink, H. J.; Neher, D.; Koster, L. J. A. Charge Transport Layers Limiting the Efficiency of

Perovskite Solar Cells: How To Optimize Conductivity, Doping, and Thickness. *ACS Appl. Energy Mater.* **2019**, *2*, 9, 6280-6287.

[19] An, Q.; Fassl, P.; Hofstetter, Y. J.; Becker-Koch, D.; Bausch, A.; Hopkinson, P. E.; Vaynzof, Y.; *Nano Energy* **2017**, 39, 400.

[20] Fagiolari, L.; Bella, F. Carbon-based materials for stable, cheaper and large-scale processable perovskite solar cells, *Energy Environ. Sci.* **2019**, 12, 3437-3472

[21] Wang, Q.; Shao, Y.; Dong, Q.; Xiao, Z.; Yuan, Y.; Huang, J. Large fill-factor bilayer iodine perovskite solar cells fabricated by a low-temperature solution-process. *Energy Environ. Sci.* **2014**, 7, 2359.

[22] An, Q.; Sun, Q.; Weu, A.; Becker-Koch, D.; Paulus, F.; Arndt, S.; Stuck, F.; Hashmi, A. S. K.; Tessler, N.; Vaynzof, Y. Enhancing the Open-Circuit Voltage of Perovskite Solar Cells by up to 120 mV using π-Extended Phosphoniumfluorene Electrolytes as Hole Blocking Layers. *Adv. Energy Mater.* **2019**, 9, 1901257.

[23] van Reenen, S.; Kemerink, M.; Snaith, H. J. Modeling Anomalous Hysteresis in Perovskite Solar Cells, *J. Phys. Chem. Lett.* **2015**, 6, 19, 3808-3814.

[24] Jacobs, D. A.; Wu, Y.; Shen, H.; Barugkin, C.; Beck, F. J.; White, T. P.; Weber, K.; Catchpole, K. R. Hysteresis phenomena in perovskite solar cells: the many and varied effects of ionic accumulation. *Phys. Chem. Chem. Phys.* **2017**, 19, 3094-3103.

[25] Richardson, G.; O'Kane, S. E. J.; Niemann, R; G.; Peltola, T. A.; Foster, J. M.; Cameron, P. J.; Walker, A. B. Can slow-moving ions explain hysteresis in the current–voltage curves of perovskite solar cells? *Energy Environ. Sci.* **2016**, 9, 1476-1485.

[26] Tessler, N.; Vaynzof, Y. Preventing Hysteresis in Perovskite Solar Cells by Undoped Charge Blocking Layers. *ACS Appl. Energy Mater.* **2018**, 1, 2, 676-683.

[27] O'Kane, S. E. J.; Richardson, G.; Pockett, A.; Niemann, R. G.; Cave, J. M.; Sakai, N.; Eperon, G. E.; Snaith, H. J.; Foster, J. M.; Cameron, P. J.; Walker, A. B. Measurement and modelling of dark current decay transients in perovskite solar cells. *J. Mater. Chem. C*, **2017**, 5, 452-462.

[28] Walter, D.; Fell, A.; Wu, Y.; Duong, T.; Barugkin, C.; Wu, N.; White, T.; Weber, K. Transient Photovoltage in Perovskite Solar Cells: Interaction of Trap-Mediated Recombination and Migration of Multiple Ionic Species. *J. Phys. Chem. C* **2018**, 122, 21, 11270-11281.

[29] Scherkar, T. S.; Momblona, C.; Gil-Escrig, L.; Bolink, H. J.; Koster, L. J. A. Improving Perovskite Solar Cells: Insights From a Validated Device Model, *Adv. Energy Mater.* **2017**, 7, 1602432.

[30] Xu, L.; Imenabadi, R. M.; Vandenberghe, W. G.; Hsu, J. W. P. Minimizing performance degradation induced by interfacial recombination in perovskite solar cells through tailoring of the transport layer electronic properties, *APL Mater.* **2018**, 6, 036104.

[31] Wu, N.; Wu, Y.; Walter, D.; Shen, H.; Duong, T.; Grant, D.; Barugkin, C.; Fu, X.; Peng, J.; White, T.; et al. Identifying the Cause of Voltage and Fill Factor Losses in Perovskite Solar Cells by Using Luminescence Measurements, Energy Tech. 2017, 5, 1827-1835.

[32] Baig, F.; Khattak, Y. H.; Mari, B.; Beg, S.; Ahmed, A.; Kahn K. Efficiency Enhancement of $CH_3NH_3SnI_3$ Solar Cells by Device Modeling. *J. Electron. Mater.* **2018**, 47, 5275-5282.

[33] Sajid, S.; Elseman, A. M.; Ji, J.; Dou, S.; Wei, S.; Huang, H.; Cui, P.; Xi, W.; Chu, L.; Li Y.; et al. Computational Study of Ternary Devices: Stable, Low-Cost, and Efficient Planar Perovskite Solar Cells. Nano Micro Lett. 2018, 10, 51.

[34] Zhou, Q.; Jiao, D.; Fu, K.; Wu, X.; Chen, Y.; Lu, J.; Yang, S.-e. Two-dimensional device modeling of $CH_3NH_3PbI_3$ based planar heterojunction perovskite solar cells. *Solar Energy* **2016**, 123, 51-56.

[35] Olyaeefar, B.; Ahmadi-Kandjani, S.; Asgari, A. Classical modelling of grain size and boundary effects in polycrystalline perovskite solar cells. *Sol. Energy Mater. Sol. Cells* **2018**, 180, 76-82.

[36] Sherkar, T. S.; Momblona, C.; Gil-Escrig, L.; Ávila, J.; Sessolo, M.; Bolink, H. J. Koster, J. A. Recombination in Perovskite Solar Cells: Significance of Grain Boundaries, Interface Traps, and Defect Ions, *ACS Energy Lett.* **2017**, 2, 5, 1214-1222.

[37] Jariwala, S.; Sun, H.; Adhyaksa, G. W. P.; Lof, A.; Muscarella, L. A.; Ehler, B.; Garnett, E. C.; Ginger, D. S. Local Crystal Misorientation Influences Non-radiative Recombination in Halide Perovskites. *Joule* **2019**, in press.

[38] Wang, F.; Bai, S.; Tress, W.; Hagfeldt, A.; Gao, F. Defects engineering for high-performance perovskite solar cells, *npj Flexible Electron.* **2018**, 2, 22.

[39] Darvishzadeh, P.; Babanezhad, M.; Ahmadi, R.; Gorji, N. E. Modeling the degradation/recovery of open-circuit voltage in perovskite and thin film solar cells. *Mater. Des.* **2017**, 113, 339-344.

[40] Darvishzadeh, P.; Redzwan, G.; Ahmadi, R.; Gorji, N. E. Modeling the degradation/recovery of short-circuit current density in perovskite and thin film photovoltaics. *Org. Electron.* **2017**, 43, 247-252.

[41] Sohrabpoor, H.; Puccetti, G.; Gorji, N. E. Modeling the degradation and recovery of perovskite solar cells. *RCS Advances* **2016**, 6, 49328-49334.

[42] Farrell, P.; Rotundo, N.; Doan, D. H.; Kantner, M.; Fuhrmann, J.; Koprucki, T., Chaper 50 Drift-Diffusion Models. In Handbook of Optoelectronic Device Modeling and Simulation, Piprek, J., Ed. CRC Press: Boca Raton, **2017**.

[43] Zhou, Y.; Long, G., Low Density of Conduction and Valence Band States Contribute to the High Open-Circuit Voltage in Perovskite Solar Cells. *J. Phys. Chem. C* **2017**, 121, 1455-1462.

[44] Anusca, I.; Balčiūnas, S.; Gemeiner, P.; Svirskas, Š.; Sanlialp, M.; Lackner, G.; Fettkenhauer, C.; Belovickis, J.; Samulionis, V.; Ivanov, M.; Dkhil, B.; Banys, J.; Shvartsman, V. V.; Lupascu, D. C., Dielectric Response: Answer to Many Questions in the Methylammonium Lead Halide Solar Cell Absorbers. *Adv. Energy Mater.* **2017**, 7, 1700600.

[45] Onoda-Yamamuro, N.; Matsuo, T.; Suga, H., Dielectric Study of $CH_3NH_3PbX_3$ (X = Cl, Br, I). *J. Phys. Chem. Solids* **1992**, 53, 935-939.

[46] Lee, J.-W.; Kim, S.-G.; Yang, J.-M.; Yang, Y.; Park, N.-G. Verification and mitigation of ion migration in perovskite solar cells, *APL Mater.* **2019**, 7, 041111.

[47] Eames, C. ; Frost, J. M.; Barnes, P. R.; O'regan, B. C.; Walsh, A.; Islam, M. S. Ionic transport in hybrid lead iodide perovskite solar cells. *Nat. Commun.* **2015**, 6, 7497.

[48] Yang, T. Y.; Gregori, G.; Pellet, N.; Grätzel, M.; Maier, J. The Significance of Ion Conduction in a Hybrid Organic–Inorganic Lead-Iodide-Based Perovskite Photosensitizer. *Angew. Chem.* **2015**, 127, 8016.

[49] Li, C.; Tscheuschner, S.; Paulus, F.; Hopkinson, P.E.; Kießling, J.; Köhler, A.; Vaynzof, Y.; Huettner, S. Iodine Migration and its Effect on Hysteresis in Perovskite Solar Cells. *Adv. Mater.* **2016**, 28, 2446.

[50] Calado, P.; Telford, A. M.; Bryant, D.; Li, X.;, Nelson, J.; O'Regan, B. C.; Barnes, P. R. F. Evidence for Ion Migration in Hybrid Perovskite Solar Cells with Minimal Hysteresis. *Nat. Commun.* **2016**, 7, 13831.

[51] Courtier, N. E.; Cave, J. M.; Foster, J. M.; Walker, A. B.; Richardson, G., How Transport Layer Properties Affect Perovskite Solar Cell Performance: Insights from a Coupled Charge Transport/Ion Migration Model. *Energy Environ. Sci.* **2019**, *12*, 396-409.

[52] Moia, D.; Gelmetti, I.; Calado, P.; Fisher, W.; Stringer, M.; Game, O.; Hu, Y.; Docampo, P.; Lidzey, D.; Palomares, E.; Nelson, J.; Barnes, P. R. F., Ionic-to-Electronic Current Amplification in Hybrid Perovskite Solar Cells: Ionically Gated Transistor-Interface Circuit Model Explains Hysteresis and Impedance of Mixed Conducting Devices. *Energy Environ. Sci.* **2019**, 12, 1296-1308.

[53] Ryu, S.; Noh, J. H.; Jeon, N. J.; Kim, Y. C.; Yang, W. S.; Seo, J.; Il Seok, S. Voltage output of efficient perovskite solar cells with high open-circuit voltage and fill factor, *Energy Environ. Sci.* **2014**, 7 , 2614-2618.

[54] Polander, L. E.; Pahner, P.; Schwarze, M.; Saalfrank, M.; Koerner, C.; Leo, K. Hole-transport material variation in fully vacuum deposited perovskite solar cells. *APL Mater.* **2014**, 2, 081503.

[55] Kulbak, M.; Cahen, D.; Hodes, G. How Important Is the Organic Part of Lead Halide Perovskite Photovoltaic Cells? Efficient $CsPbBr_3$ Cells. *J. Phys. Chem. Lett.* **2015**, 6 , 2452-2456.

[56] Belisle, R. A.; Jain, P.; Prasanna, R.; Leijtens, T.; McGehee, M. D. Minimal Effect of the Hole-Transport Material Ionization Potential on the Open-Circuit Voltage of Perovskite Solar Cells, *ACS Energy Lett.* **2016**, 1, 556-560.

[57] Dänekamp, B.; Droseros, N.; Tsokkou, D.; Brehm, V.; Boix, P. P.; Sessolo, M.; Banerji, N.; Bolink, H. J. Influence of hole transport material ionization energy on the performance of perovskite solar cells. *J. Mater. Chem. C* **2019**, 7, 523-527.

[58] Chen, S.; Hou, Y.; Chen, H.; Richter, M.; Guo, F.; Kahmann, S.; Tang, X.; Stubhan, T.; Zhang, H.; Li, N.; Gasparini, N.; Ramirez Quiroz, C. O.; Khanzada, L. S.; Matt, G. J.; Osvet, A.;

Brabec, C.J. Exploring the Limiting Open-Circuit Voltage and the Voltage Loss Mechanism in Planar CH$_3$NH$_3$PbBr$_3$ Perovskite Solar Cells. *Adv. Energy Mater.* **2016** 6, 1600132.

[59] Ravishankar, S.; Gharibzadeh, S.; Roldán-Carmona, C.; Grancini, G.; Lee, Y.; Ralaiarisoa, M.; Asiri, A. M.; Koch, N.; Bisquert, J.; Nazeeruddin, M. K. Influence of Charge Transport Layers on Open-Circuit Voltage and Hysteresis in Perovskite Solar Cells. *Joule* **2018**, 2, 788-798.

[60] Shao, S.; Chen, Z.; Fang, H.-H.; ten Brink, G. H.; Bartesaghi, D.; Adjokatse, S.; Koster, L. J. A.; Kooi, B. J.; Facchetti, A.; Loi, M. A. N-type polymers as electron extraction layers inhybrid perovskite solar cells with improved ambient stability. *J. Mater. Chem. A* **2016**, 4, 2419.

[61] Le Corre, V. M.; Stolterfoht, M.; Perdigon Toro, L.; Feuerstein, M.; Wolff, C.; Gil-Escrig, L.; Bolink, H. J.; Neher, D.; Koster, L. J. A. Charge transport layers limiting the efficiency of perovskite solar cells: how to optimize conductivity, doping and thickness. *ACS Appl. Energy Mater.* **2019**, 2, 6280-6287.

[62] Wolff, C. M.; Caprioglio, P.; Stolterfoht, M.; Neher, D. Nonradiative Recombination in Perovskite Solar Cells: The Role of Interfaces. *Adv. Mater*. **2019**, DOI: 10.1002/adma.201902762.

[63] Shao, Y.; Yuan, Y.; Huang, J. Correlation of energy disorder and open-circuit voltage in hybrid perovskite solar cells, *Nat. Energy* **2016**, 1, 15001.

[64] Bi, C.; Wang, Q.; Shao, Y.; Yuan, Y.; Xiao, Z.; Huang, J. Non-wetting surface-driven high-aspect-ratio crystalline grain growth for efficient hybrid perovskite solar cells. *Nat. Commun*. **2015**, 6, 7747.

[65] Zhang, L.; Wu, J.; Li, D.; Li, W.; Meng, Q.; Bo, Z. Ladder-like conjugated polymers used as hole-transporting materials for high-efficiency perovskite solar cells. *J. Mater. Chem. A* **2019**, 7, 14473-14477.

[66] Pazokiand, M.; Edvinsson, T. Metal replacement in perovskite solar cellmaterials: chemical bonding effects andoptoelectronic properties. *Sustain. Energy Fuels* **2018**, 2, 1430-1445.

[67] Zhou, Y; Long, G. Low Density of Conduction and Valence Band States Contribute to the High Open-Circuit Voltage in Perovskite Solar Cells. *J. Phys. Chem. C* **2017**, 121, 1455-1462.

[68] Endres, J.; Egger, D. A.; Kulbak, M.; Kerner, R. A.; Zhao, L.;Silver, S. H.; Hodes, G.; Rand, B. P.; Cahen, D.; Kronik, L.; et al. Valence and Conduction Band Densities of States of Metal Halide Perovskites: A Combined Experimental-Theoretical Study. *J. Phys. Chem. Lett.* **2016**, 7, 2722-2729.

[69] Roichman, Y.; Tessler, N., Generalized Einstein Relation for Disordered Semiconductors - Implications for Device Performance. *Appl. Phys. Lett.* **2002**, 80, 1948-1950.

[70] Stranks, S. D. *ACS Energy Lett.* **2017**, 2, 1515.

[71] Neukom, M. T.; Schiller, A.; Züfle, S.; Knapp, E.; Ávila, J.; Pérez-del-Rey, D.; Dreessen, C.; Zanoni, K. P. S.; Sessolo, M.; Bolink, H. J.; Ruhstaller, B., Consistent Device Simulation Model Describing Perovskite Solar Cells in Steady-State, Transient, and Frequency Domain. *ACS Appl. Mater. Interfaces* **2019**, 11, 23320-23328.

[72] Rozenberg, M. J.; Inoue, I. H.; Sánchez, M. J., Nonvolatile Memory with Multilevel Switching: A Basic Model. *Phys. Rev. Lett.* **2004**, 92, 178302.

[73] Butscher, J. F.; Intorp, S.; Kress, J.; An, Q.; Hofstetter, Y. J.; Hippchen, N.; Paulus, F.; Bunz, U.H.F.; Tessler, N.; Vaynzof, Y. Enhancing the Open-Circuit Voltage of Perovskite Solar Cells by Embedding Molecular Dipoles within Their Hole-Blocking Layer, *ACS Appl. Mater. Interfaces* **2019***,* in press https://doi.org/10.1021/acsami.9b18757.

[74] Lee, J.-H.; Kim, J.; Kim, G.; Shin, D.; Jeong, S. Y.; Lee, J.; Hong, S.; Choi, J. W.; Lee, C.-L.; Kim, H.; Yi, Y.; Lee, K. Introducing paired electric dipole layers for efficient and reproducible perovskite solar cells, *Energy Environ. Sci.* **2018**, 11, 1742-1751.

[75] Butscher, J. F.; Sun, Q.; Wu, Y.; Stuck, F.; Hoffmann, M.; Dreuw, A.; Paulus, F.; Hashmi, A. S. K.; Tessler, N.; Vaynzof, Y. Dipolar Hole-Blocking Layers for Inverted Perovskite Solar Cells: Effects of Aggregation and Electron Transport Levels, J. Phys. Mater. **2020**, 3, 025002.

[76] Egger, D. A.; Bera, A.; Cahen, D.; Hodes, G.; Kirchartz, T.; Kronik, L.; Lovrincic, R.; Rappe, A. M.; Reichman, D. R.; Yaffe, O., What Remains Unexplained About the Properties of Halide Perovskites? *Adv. Mater.* **2018**, 30, 11.

[77] Lopez-Varo, P.; Jiménez-Tejada Juan, A.; García-Rosell, M.; Ravishankar, S.; Garcia-Belmonte, G.; Bisquert, J.; Almora, O., Device Physics of Hybrid Perovskite Solar Cells: Theory and Experiment. *Adv. Energy Mater.* **2018**, 8, 1702772.

[78] Li, C.; Guerrero, A.; Zhong, Y.; Gräser, A.; Luna, C. A. M.; Köhler, J.; Bisquert, J.; Hildner, R.; Huettner, S., Real-Time Observation of Iodide Ion Migration in Methylammonium Lead Halide Perovskites. *Small* **2017**, 13, 1701711

[79] deQuilettes, D.W.; Zhang, W.; Burlakov, V. M.; Graham, D. J.; Leijtens, T.; Osherov, A.; Bulović, V.; Snaith, H. J.; Ginger, D. S.; Stranks, S. D. Photo-induced halide redistribution in organic–inorganic perovskite films. *Nat. Commun.* **2016**, 7, 11683.

[80] Walsh, A.; Stranks, S. D., Taking Control of Ion Transport in Halide Perovskite Solar Cells. *ACS Energy Lett.* **2018**, 3, 1983-1990.

[81] Xu, L.; Molaei Imenabadi, R.; Vandenberghe, W. G.; Hsu, J. W. P., Minimizing Performance Degradation Induced by Interfacial Recombination in Perovskite Solar Cells through Tailoring of the Transport Layer Electronic Properties. *APL Materials* **2018**, 6, 036104.

[82] Galatopoulos, F.; Papadas, I. T.; Armatas, G. S.; Choulis, S. A., Long Thermal Stability of Inverted Perovskite Photovoltaics Incorporating Fullerene-Based Diffusion Blocking Layer. *Adv. Mater. Interfaces* **2018**, 5, 1800280.

[83] Wu, S.; Chen. R.; Zhang, S.; Babu, B. H.; Yue, Y.; Zhu, H.; Yang, Z.; Chen, C.; Chen, W.; Huang, Y.; Fang, S.; Liu, T.; Han, L,; Chen, W.; A chemically inert bismuth interlayer enhances long-term stability of inverted perovskite solar cells. *Nat. Commun.* **2019**, 10, 1161.

[84] Papadas, I.T.; Galatopoulos, F.; Armatas, G.S.; Tessler, N.; Choulis, S.A. Nanoparticulate Metal Oxide Top Electrode Interface Modification Improves the Thermal Stability of Inverted Perovskite Photovoltaics. *Nanomaterials* **2019**, *9*, 1616.

[85] Nan, G.; Zhang, X.; Lu, G. Self-Healing of Photocurrent Degradation in Perovskite Solar Cells: The Role of Defect-Trapped Excitons. *J. Phys. Chem. Lett.* **2019**, 10, 24, 7774-7780.

[86] Ceratti, D. R.; Rakita, Y.; Cremonesi, L.; Tenne, R.; Kalchenko, V.; Elbaum, M.; Oron, D.; Potenza, M. A. C.; Hodes, G.; Cahen, D. Self-healing inside APbBr$_3$ Halide Perovskite Crystals. *Adv. Mater.* **2019**, 30 (10), 1706273.